 \documentclass[preprint2]{aastex}

\shorttitle{Silica-Rich Bright Debris Disk around HD~15407A}
\shortauthors{Fujiwara et al.}

\begin{document}

\title{Silica-Rich Bright Debris Disk around HD~15407A}


\author{
Hideaki~Fujiwara\altaffilmark{1}, 
Takashi~Onaka\altaffilmark{2}, 
Takuya~Yamashita\altaffilmark{3}, 
Daisuke~Ishihara\altaffilmark{4}, 
Hirokazu~Kataza\altaffilmark{5}, 
Misato~Fukagawa\altaffilmark{6}, 
Yoichi~Takeda\altaffilmark{3}, and 
Hiroshi~Murakami\altaffilmark{5} 
}


\altaffiltext{1}{Subaru Telescope, National Astronomical Observatory of Japan, 650 North A'ohoku Pl., Hilo, HI 96720, U.S.A;  hideaki@naoj.org}
\altaffiltext{2}{Department of Astronomy, School of Science, University of Tokyo, Bunkyo-ku, Tokyo 113-0033, Japan}
\altaffiltext{3}{National Astronomical Observatory of Japan, 2-21-1 Osawa, Mitaka, Tokyo 181-0015, Japan}
\altaffiltext{4}{Nagoya University, Furo-cho, Chikusa-ku, Nagoya 464-8602, Japan}
\altaffiltext{5}{Institute of
Space and Astronautical Science, Japan Aerospace Exploration Agency, 
3-1-1 Yoshinodai, Chuo-ku, Sagamihara, Kanagawa 252-5210, Japan}
\altaffiltext{6}{Graduate School of Science, Osaka University, 1-1 Machikaneyama, 
Toyonaka 560-0043, Osaka, Japan}


\begin{abstract}
We report an intriguing debris disk towards the F3V star HD~15407A, 
in which an extremely large amount of warm fine dust ($\sim 10^{-7}M_\oplus$) is detected. 
The dust temperature is derived as $\sim$ 500--600~K and the location of the debris dust 
is estimated as 0.6--1.0~AU from the central star, a terrestrial planet region. 
The fractional luminosity of the debris disk is $\sim 0.005$, which is much larger 
than those predicted by steady-state models of the debris disk produced by planetesimal collisions. 
The mid-infrared spectrum obtained by {\it Spitzer} indicates the presence of 
abundant $\micron$-sized silica dust, 
suggesting that the dust comes from the surface layer of differentiated large rocky bodies 
and might be trapped around the star. 
\end{abstract}




\section{Introduction}

As we observed signs of collisions of comets onto Jupiter in 1994 and 2009 \citep[e.g.][]{hammel10} 
and collisions between asteroids \citep[e.g.][]{ishiguro11a}, 
collisions between small bodies and/or planets are common in the Solar System. 
In the younger Solar System, collisions occurred more frequently 
and were one of the most important processes in its formation and evolution \citep{kenyon04}. 
Especially, hypervelocity collisions have direct links to catastrophic phenomena on planets, 
such as craters on the planets, moon formation through the giant impact, 
and mass extinction on the Earth \citep[e.g.][]{bottke07}. 
In the extrasolar system, products of collisions among planetesimals around young stars are 
observed as debris disks \citep{wyatt08}. 

HD~15407A (HIP~11696) is an F3V main-sequence star, whose distance from the Sun is 55~pc \citep{vanleeuwen07}. 
The age, the effective temperature, and the metallicity of the star have been estimated as $2.1 \pm 0.3$~Gyr, $6350$~K, 
and [Fe/H] $= +0.08$, respectively, by Geneva-Copenhagen Survey \citep{holmberg09}. 
The parameters are confirmed by high-dispersion spectra obtained with Okayama Astrophysical Observatory (OAO)/HIDES 
and the comparison of the position of HD~15407A in the H-R diagram to theoretical tracks and isochrones based on the method described in \cite{takeda07}.
On the other hand, a recent paper suggests 
a much younger age ($80^{+40}_{-20}$~Myr) from the lithium 6710 \AA \ absorption and the motion in the Galaxy \citep{melis10}. 
HD~15407A is suggested as a star in double system with the nearby K2V star HD~15407B (HIP~11692) in the SIBMAD database. 
The separation between the two stars is $21\farcs2$ and the projected distance is $1170$~AU at the distance of 55~pc.

Large excess over the expected stellar photospheric emission 
at 12 and 60$~\micron$, whose color temperature is estimated as 750~K, 
has been suggested by {\it IRAS} observations \citep{oudmaijer92}. 
The infrared excess was confirmed by the recent {\it AKARI} \citep{murakami07} 
mid-infrared (MIR) All-Sky Survey at $9$ and $18~\micron$ \citep{ishihara10}, 
suggesting the presence of a warm dust disk. 

In this {\it Letter}, we present new MIR photometry of HD~15407A obtained with {\it AKARI} 
as well as a MIR low resolution spectrum obtained with {\it Spitzer}. 
We report the detection of abundant silica dust together with amorphous silicate dust towards the star 
and discuss the nature of the debris disk around HD~15407A.

\section{Observations and Data Reduction}

\subsection{{\it AKARI}/IRC all-sky survey}

The S9W (9$~\micron$) and L18W (18$~\micron$) images of HD~15407A were taken 
with the Infrared Camera \citep[IRC;][]{onaka07} onboard {\it AKARI} 
as part of the All-Sky Survey observations \citep{ishihara10}. 
The star was observed on Aug 18--17, 2006 and Feb 15--16, 2007 (UT). 
The fluxes at the two periods agree with each other within the uncertainty, 
indicating no appreciable variations in the flux. 
HD~15407A was selected as a probable candidate of the debris disk with 18$~\micron$ excess 
from the early products of the {\it AKARI}/IRC All-Sky Survey. 
The flux densities are determined as $0.960 \pm 0.031$~Jy and $0.497 \pm 0.021$~Jy (AKARI-IRC-V1~J0230506+553254), 
in the S9W and L18W bands, respectively, in the {\it AKARI}/IRC Point Source Catalog \citep{ishihara10}. 
Although the {\it AKARI} flux densities of HD~15407A might be contaminated by HD~15407B 
due to the large beam sizes ($5\farcs5$ and $5\farcs7$ for the S9W and L18W bands, respectively), 
the fractional flux densities of the contamination are estimated 
to be $< 20$\% in both bands and thus the detection of the excess is secure.

\subsection{{\it Spitzer}/IRS observations}

Follow-up spectroscopic observations of HD~15407A were made with the Infrared Spectrograph \citep[IRS;][]{houck04} 
onboard {\it Spitzer} on Oct 9, 2008 (AOR ID\# 26122496). 
All the four low-resolution modules, Short-Low 2 (5.2--7.7$~\micron$) 
and Short-Low 1 (7.4--14.5$~\micron$), Long-Low 2 (14.0--21.3$~\micron$) and Long-Low 1 (19.5--38.0$~\micron$), 
were used to obtain a full 5--35$~\micron$ low-resolution ($\lambda/\Delta \lambda \sim 100$) spectrum. 
We used the pipeline-processed (S18.1) Basic Calibrated Data products 
for the target and analyzed the sky-subtracted extracted 1D spectral data 
since the target is a point source in a relatively empty field. 
The wavelength calibration is as good as 0.1$~\micron$ along the dispersal direction. 
For the Long-Low 1 spectra, we use the data only for $\lambda < 35~\micron$ because the noise becomes 
large at $\lambda > 35~\micron$, whose spectral range is not critical for the present analysis. 
We adopt the flux densities of the pipeline-processed products, for which the absolute accuracy 
in the Short-Low and Long-Low spectra is better than 10\%. 
The {\it Spitzer} spectrum does not include the contribution of HD~15407B
since HD~15407A and HD~15407B are clearly separated in the peak-up image of the IRS.




\section{Results}

\subsection{Spectral Energy Distribution}

The observed flux densities of HD~15407A in various bands are shown in Table~\ref{photometry}. 
The photospheric flux densities of HD~15407A are estimated from the Kurucz model \citep{kurucz92} 
with the effective temperature of $6500$~K and the surface gravity of $\log g=+4.0$ 
fitted to the 2MASS $K_{\rm S}$-band photometry of the star. 
The resultant photospheric spectrum fitted with the $K_{\rm S}$-band is in agreement with the 2MASS $J$- and $H$-band data. 
Therefore the extinction to HD~15407A is negligible in the infrared. 
The extinction in the optical region is also negligible since the observed $B-V$ color 
of the star is consistent with the intrinsic color of F2--5V stars ($B-V=0.35$ for F2, and 0.44 for F5 stars). 

\begin{figure}
\epsscale{1.2}
\plotone{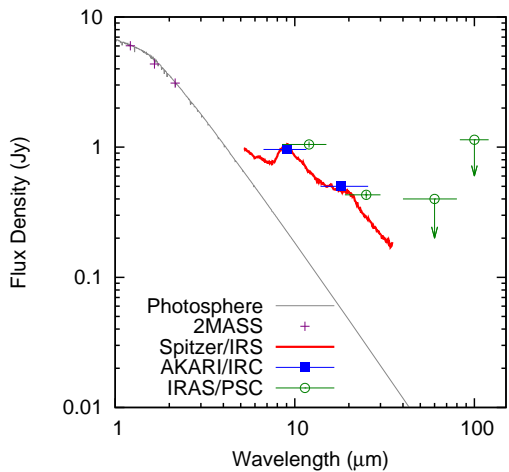}
\caption{
SED of HD~15407A. The blue filled squares, green open circles, and purple crosses indicate the photometric data obtained with {\it AKARI}/IRC, {\it IRAS}, and 2MASS, respectively. The horizontal bars of the photometric data show the bandwidths. The red and gray lines indicate the {\it Spitzer}/IRS spectrum and the photospheric contribution of F3 star \citep{kurucz92}, respectively. 
\label{SED_hd15407}}
\end{figure}

The obtained spectral energy distribution (SED) of the star 
compiled with the {\it Spitzer}/IRS spectrum and the {\it AKARI} and {\it IRAS} photometry is shown in Figure~\ref{SED_hd15407} 
together with the expected photospheric emission. 
Significant excess emission at wavelengths longer than 5$~\micron$ is clearly seen 
and the detected flux densities at 9 and 18$~\micron$ are 5 and 10 times larger 
than the photosphere, unambiguously indicating the presence of a warm and bright debris disk.

\subsection{Dust Features}

To examine the excess emission in detail, 
we subtract the estimated photospheric emission from the observed IRS spectrum (Figure~\ref{fit_fused}). 
The photosphere-subtracted spectrum shows two significant emission features centered at around 9 and 20$~\micron$. 
A weak feature at 16$~\micron$ is also seen. 
A ubiquitous dust species family, silicate, which has spectral features around $\sim 10~\micron$ 
and $\sim 20~\micron$, 
can be considered as a main carrier of the observed features at the first sight. 
However, sub-$\micron$-sized amorphous olivine and pyroxene commonly seen in celestial objects 
have features around 9.3--9.7$~\micron$ and $18.0~\micron$, 
which cannot reproduce the observed spectrum completely. 
As an alternative carrier of the observed features around around 9 and 20$~\micron$, 
we consider amorphous silica (silicon dioxide), which also has Si-O and O-Si-O modes and shows broad features 
at similar wavelengths to amorphous silicate.

\subsection{Spectral Fitting}

To estimate the dust temperature and the species, we perform a fit with an SED model that consists of blackbody emission, silicate and silica emission. The model flux density is given by 
\begin{eqnarray*}
F_{{\rm exc},\nu}^{\rm mod}(\lambda) &=& \Omega_{\rm BB} B_\nu(\lambda, T_{\rm BB})\\
&+& a_{\rm silicate} \kappa_{\rm silicate} B_\nu(\lambda, T_{\rm silicate}) \\
&+& a_{\rm silica} \kappa_{\rm silica} B_\nu(\lambda, T_{\rm silica}), 
\end{eqnarray*}
where $\Omega_{\rm BB}$ is the solid angle of a blackbody representing the continuum emission from large dust ($\gtrsim 10~\micron$ in size), $B_\nu(\lambda, T)$ is the Planck function with the temperatures $T$, $\kappa(\lambda)$ is the mass absorption coefficients of the dust, and $a$'s are the scaling factors proportional to the dust mass. 
For silicate, we choose one from eight mass absorption coefficients (amorphous olivine or pyroxene with the size of 0.1, 1.0, 1.5, or 2.0$~\micron$) computed from the optical constants of MgFeSiO$_4$ and Mg$_{0.5}$Fe$_{0.5}$SiO$_{3}$ \citep{dorschner95} based the Mie theory \citep{bohren83}. We take the specific mass densities of both olivine and pyroxene as 3.3 g~cm$^{-3}$. For silica, we choose one from fused quartz \citep{koike89} or annealed silica \citep{fabian00}. Fused quartz is an amorphous (glassy) polymorph of silica and shows broad features in the MIR. Annealed silica may contain cristobalite and tridymite, polymorphs of crystalline silica, and shows relatively sharp features. We take the mass absorption coefficients of annealed silica and compute that of fused quartz for the shape of continuous ellipsoidal distribution \citep{fabian01} from the optical constants measured in laboratories \citep{koike89}. 
We search for parameters minimizing the reduced $\chi^2_\nu$  
using the data between 5 and 35$~\micron$ with respect to each of $\Omega_{\rm BB}$ and $a$'s.

\begin{figure}
\epsscale{1.0}
\plotone{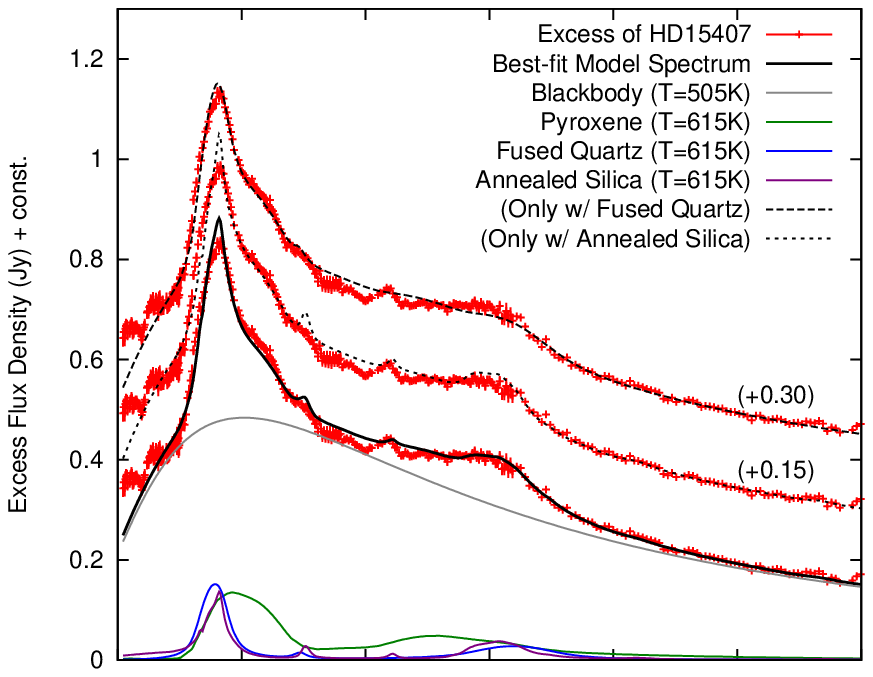}
\vspace{-0.6cm}
\plotone{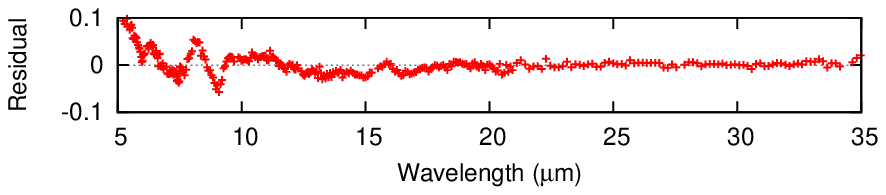}
\caption{
Debris disk spectrum of HD~15407A (red crosses) and the best-fit model spectrum (solid black line) consisting of blackbody dust (gray), 1.5$~\micron$-sized amorphous pyroxene (green), fused quartz (blue), and annealed silica (purple). The debris disk spectrum is derived by subtraction of the estimated photospheric contribution from the observed IRS spectrum. 
The residual spectra subtracted by the best-fit model is shown below the fit result. 
The model spectrum with fused quarts or annealed silica only as silica dust is also shown shown in the main panel as dashed and dotted black lines, respectively. 
\label{fit_fused}}
\end{figure}

At first, we find that a combination of blackbody, 1.5$~\micron$-sized amorphous pyroxene and fused quartz 
provides the best fit (Figure~\ref{fit_fused}). The second best combination is blackbody, 1.5$~\micron$-sized amorphous pyroxene, and annealed silica. The best-fit model spectra and parameters of these combinations are shown in Figure~\ref{fit_fused} and Table~\ref{fitparameter}, respectively. 
The models with the two combinations of dust components seem fairly good, 
suggesting that it is a robust conclusion that silica dust is one of the major dust species of debris dust 
together with amorphous pyroxene dust around HD~15407A.
The model spectrum with annealed silica explains the observed small feature at $\lambda = 16~\micron$ 
while it produces a sharp extra emission at $\lambda = 12.6~\micron$, which is not observed. 
Since fused quartz and annealed silica 
have similar but slightly different band features, 
we perform a tuned fit assuming that the silica component is a mixture of fused quartz and annealed silica, 
and that the silicate is 1.5$~\micron$-sized amorphous pyroxene with $T_{\rm silicate}=T_{\rm silica}$. 
The best-fit model spectrum and parameters in this dust combination 
are shown in Figure~\ref{fit_fused} and Table~\ref{fitparameter}, respectively. 
Since this model provides the least $\chi^2_\nu$ among the models considered here, 
we adopt this model ($T_{\rm BB}=505$~K and $T_{\rm silicate}=T_{\rm silica}=615$~K) in the following discussion. 
The masses of amorphous pyroxene and silica dust around HD~15407A are derived 
as $4.2 \times 10^{17}$ and $2.5 \times 10^{17}$~kg, respectively, assuming that the dust emission is optically thin. 
The derived total mass of the fine dust in the warm debris disk is $6.8 \times 10^{17}$~kg $\sim 10^{-7}M_\oplus$. 
This value does not include 
the mass of the blackbody dust, which is supposed to come from large-sized ($\gtrsim 10~\micron$ in size) grains and rubbles.

\section{Discussion}
\subsection{Radial Distribution of Dust}

We estimate the distances from the central star of the blackbody and the 1.5$~\micron$-sized amorphous pyroxene dust 
from the dust temperatures and emissivities assuming that the grains are in radiative equilibrium around a F3V star with the luminosity of $3.9L_\odot$. 
The derived distance for the blackbody and the pyroxene dust is 0.6 and 1.0~AU, respectively, 
which corresponds to a terrestrial (rocky) planet region. 
The silica dust must contact with other dust species to keep its temperature high since it is transparent in the visible wavelength and the stellar radiation solely cannot 
heat silica grains to the observed temperature. 
Thus, although the distance of silica from the star in radiative equilibrium is $<0.1$~AU, the silica dust is considered to be located at a region similar to those of pyroxene and blackbody dust. 
Only upper limits are available for the far-infrared (FIR) flux densities from {\it IRAS} observations and the extension of the debris disk cannot be estimated. 
High spatial resolution observations and FIR--radio observations are required for further examination of the radial distribution of dust around the star.

\subsection{Fractional Luminosity} 

The fractional luminosity ($f_{\rm obs}=L_{\rm dust}/L_{\rm star}$) of the warm debris disk is estimated as $\sim 0.005$ from the fitting of the SED. 
Figure 3 plots the maximum fractional luminosity $f_{\rm max}$ predicted by a simple model of the steady-state evolution of debris disks produced by collisions \citep{wyatt07} 
with the stellar parameters of HD~15407A and the derived dust disk radii ($R_{\rm dust} = 0.6$ and 1.0~AU for blackbody and pyroxene dust, respectively). 
The ratio of $f_{\rm obs}/f_{\rm max}$ of the debris disk around HD~15407A is larger than $10^5$ if we assume 2.1~Gyr as the age of HD~15407A as suggested by \cite{holmberg09}. 
The observed fractional luminosity of HD~15407A system cannot be accounted for by a simple steady-state model. 
Even when we assume the age of $80^{+40}_{-20}$~Myr suggested by \cite{melis10}, $f_{\rm obs}/f_{\rm max}$ is about $10^4$ 
and significant enhancement of $f_{\rm obs}$ compared to $f_{\rm max}$ is secure. 
To date several warm debris disks are known, whose fractional luminosities are much larger than the steady-state model and in which transient events are suggested to be responsible; 
e.g.\ \cite{wyatt07} estimated $f_{\rm obs}/f_{\rm max}=10^3$--$10^4$ for BD~+20~307, HD~72905, $\eta$~Corvi, and HD~69830, which possess warm dust at $0.2$--$2$~AU.
HD~15407A is one of the highest-$f_{\rm obs}/f_{\rm max}$ debris disks among these and seems a non-typical warm debris disk.  
The disk is not a protoplanetary one since H$\beta$, H$\gamma$, and H$\delta$ lines are seen in absorption, not in emission, 
in the OAO/HIDES optical spectrum of the star, which suggests the absence of gas accretion around the star. 

HD~15407 is a possible binary system and HD~15407B might disturb circumstellar material around HD~15407A dynamically. 
At present there is no theoretical model of debris disk evolution for a binary system in terms of fractional luminosity. 
However, the study of the stability zone in a binary system by \cite{holman99} suggests that material at $\lesssim 400$ and $\lesssim 50$~AU from HD~15407A is dynamically stable for cases 
where the eccentricity of the system is 0.0 and 0.8, respectively.

The radiation pressure should blow out $\micron$-sized small grains, which show prominent feature in the MIR, 
in the vicinity of a F3 star in a short ($<100$~years) time scale. 
It cannot be concluded only from available data whether or not HD~15407A shows a time variation of its feature emission. 
Time variation in the continuum flux level of the excess is not seen in the observations after {\it IRAS}, 
which can be accounted for since grains responsible for the continuum excess are large ($>10~\micron$) and thus have long ($>1000$~years) lifetime.
Monitoring observations of MIR spectrum would be important to examine the actual lifetime of small dust around HD~15407A. 
If dust features towards HD~15047A do not change over decades, mechanisms that retain the fine grains near the star should be playing a role in HD~15407A, 
such as shepherding of dust around a planet as in Saturn \citep{spahn89} and Uranus \citep{murray90} 
or dust capture in a resonance as in the circumsolar dust ring \citep{reach95}, which are in effect in the Solar System. 
Alternatively the fine dust may be produced violently and continuously over decades around HD~15407A. 
It should be noted that the \cite{wyatt07} model does not assume the presence of planet and trapping dust in a resonance in a planetary system.
There is no evidence of a planet orbiting around HD~15407A at present. 
But if trapping of dust takes place efficiently, the predicted fractional luminosity from the model may be different than in debris disks without planets.


\subsection{Silica in Debris Disks} 

The mineralogical characteristics of the detected dust provide a clue for the origin of the debris dust. 
In the Solar System, silica is one of the most abundant minerals in the Earth's crust. 
The presence of amorphous silica is also suggested from the analysis of the Comet 81P/Wild 2 dust sample retrieved by the STARDUST mission \citep{mikouchi07}. 
The presence of annealed silica dust in the protoplanetary disk around several T Tauri stars (TTSs) has been indicated in their {\it Spitzer} MIR spectra \citep{sargent09}. 
However, silica is not a major constituent of the interstellar medium since the expected 9$~\micron$ feature is not seen in the interstellar medium towards the Galactic Center \citep{kemper04}. 

The presence of silica-like dust has also been suggested in debris disks.
An almost equivalent amount of $\micron$-sized amorphous pyroxene and fused quartz are detected towards HD~15407A in addition to the blackbody component. 
Ground-based observations of $8$--$13~\micron$ spectroscopy indicate 
the presence of the $9~\micron$-peaked broad feature which is attributable to amorphous silica
around the 100~Myr-old G0 field star HD~23514 \citep{rhee08}. 
{\it Spitzer}/IRS spectroscopy reveals the presence of abundant amorphous silica dust around the A0V star HD~172555 whose age is estimated as $\sim12$ Myr \citep{lisse09}. 
Detections of silica among protoplanetary disks around TTSs, debris disks, and the Solar System suggest that silica might be ubiquitously present through the planet formation processes. 
Debris disk stars with silica dust share a common habit that they show excesses emission in the near-infrared ($\lambda \lesssim 5~\micron$), suggesting the presence of very warm dust in the vicinity of the stars. 
The presence of very warm dust might be important for silica grains, which are almost transparent in the optical wavelength 
and thus needed to be heated by contact with other dust species thermally and emit in the MIR. 

It is widely accepted that debris dust is produced from collisions of planetesimals. 
Therefore, if a large amount of silica-rich planetesimals around HD~15407A collide with each other actively, the presence of abundant silica dust around HD~15407A could be accounted for. 
Although the origin of silica-rich planetesimals is still an open question, a breakup of a large differentiated rocky body may be a likely possibility. 
A similar mechanism has been proposed as an origin of the enstatite-rich bright debris disk around HD~165014 \citep{fujiwara10}. 

As a probable source of silica dust around HD~172555, in which abundant SiO gas is also detected, giant hypervelocity ($>10$~km s$^{-1}$) impact between large rocky planetesimals is suggested by \cite{lisse09}. 
It may be similar to the one that formed the Moon (giant impact) or the one that stripped the surface crustal material off of Mercury's surface.
The comparable silica-to-silicate ratio in $\micron$-sized small dust around HD~15407A suggests a similar origin of the silica dust with that of HD~172555. 
This also seems harmonic with the large fractional luminosity of HD~15407A's debris, which may be connected with a transient event. 
However, the presence of SiO gas around HD~15407A is not confirmed in its {\it Spitzer}/IRS spectrum. 
Assuming that SiO gas is really absent around HD~15407A, 
a substantial time \citep[$\sim 10^3$--$10^4$~yr;][]{pahlevan07} should have elapsed after the impact and SiO gas should have disappeared due to the process of re-condensation 
even if a hypervelocity impact might have occurred and have produced silica-rich dust around HD~15407A. 
Taking account of the short lifetime of $\micron$-sized dust, dust trapping and/or continuous dust production by collisional grinding of the large rubbles are required after the possible impact. 
Further theoretical studies in addition to observations with high spatial resolution and wide wavelength coverage 
will provide hints on the possible linkage between the large infrared excess, the dynamical evolution, and the possible transient event of the intriguing debris disk around HD~15407A.

\acknowledgments

This research is based on observations with {\it AKARI} and {\it Spitzer}. 
We thank the anonymous referee, C.\ Koike, H.\ Chihara, H.\ Mutschke, L.B.F.M.\ Waters, A.\ Takigawa, H.\ Nagahara, S.\ Tachibana, 
and C.M.\ Lisse for their useful comments to improve the manuscript. This work was supported by KAKENHI (23103002).

{\it Facilities:} \facility{{\it AKARI} (ISAS/JAXA)}, \facility{{\it Spitzer} (NASA)}.



\begin{table}
\begin{center}
\caption{Infrared Photometry of HD~15407A.\label{photometry}}
\begin{tabular}{ccccc}
\tableline\tableline
$\lambda$   & $F_\nu$       & Instrument & Photosphere\tablenotemark{a} & Significance, $\chi$\tablenotemark{b} \\
($\micron$) & (Jy)          &            & (Jy)                         & \\
\tableline
9           & $0.96 \pm 0.03 $ & {\it AKARI}/IRC & 0.22  & 25 \\
12          & $1.05 \pm 0.06 $ & {\it IRAS}      & 0.12  & 16 \\
18          & $0.50 \pm 0.02 $ & {\it AKARI}/IRC & 0.06  & 22 \\
25          & $0.43 \pm 0.03 $ & {\it IRAS}      & 0.03  & 13 \\
60          & $<0.40$          & {\it IRAS}      & 0.005 &  \nodata \\
100         & $<1.14$          & {\it IRAS}      & 0.002 &  \nodata \\
\tableline
\end{tabular}
\tablenotetext{a}{From Kurucz model to fitted to 2MASS $JHK_{\rm s}$-bands data.}
\tablenotetext{b}{$\chi = ({\rm Observed} - {\rm Kurucz}) / {\rm noise}$.}
\end{center}
\end{table}



\begin{table}
\begin{center}
\caption{Best-fit Parameters of Spectral Fitting\tablenotemark{a}.\label{fitparameter}}
\begin{tabular}{ccccccccc}
\tableline\tableline
                 & \multicolumn{2}{c}{Blackbody}     & \multicolumn{2}{c}{Silicate}                & \multicolumn{2}{c}{Silica}              & \\
\cline{2-3} \cline{4-5} \cline{6-7}
                 & $T_{\rm BB} $ & $\Omega_{\rm BB}$ & $T_{\rm silicate} $ & Mass                  & $T_{\rm silica} $ & Mass                & $\chi^2_\nu$ \\
Silica Component & (K)           & ($10^{-16}$~Str)  & (K)                 & ($10^{17}$~kg)        & (K)               & ($10^{17}$~kg)      &              \\
\tableline
Fused$+$Annealed & 505           & 2.3               & 615\tablenotemark{b}& 4.3                   & 615\tablenotemark{b}& $1.3+1.2$           & 5.4          \\
\tableline
Fused Quartz     & 505           & 2.3               & 725                 & 3.0                   & 605               & 2.2                 & 6.0          \\
Annealed Silica  & 515           & 2.4               & 815                 & 2.5                   & 600               & 2.6                 & 6.4          \\
\tableline
\end{tabular}
\tablenotetext{a}{The number of free parameters is 6 for all of the listed models.}
\tablenotetext{b}{Assumed $T_{\rm silicate}=T_{\rm silica}$.}
\end{center}
\end{table}

 \clearpage


\begin{figure}
\epsscale{1.0}
\plotone{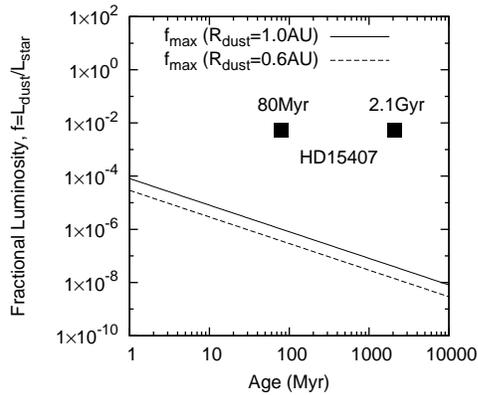}
\caption{
Fractional luminosity of the debris disk around HD~15407A. The theoretical maximum fractional luminosities of planetesimal belts at $R_{\rm dust}=0.6$ (dashed line) and 1.0~AU (solid line) around an F3V star are calculated by a model of the steady-state evolution of a debris disk produced by collisions \citep{wyatt07} with the fixed model parameters (belt width: $dr/r = 0.5$; planetesimal strength: $Q^*_D = 200$~J/kg; planetesimal eccentricity: $e = 0.05$; diameter of the largest planetesimal in cascade: $D_c = 2000$~km). 
The observed fractional luminosity towards HD~15407A (filled squares), 
which is shown by assuming the age of 2.1~Gyr \citep{holmberg09} and 80~Myr \citep{melis10}, 
is $10^4$--$10^5$ times larger than that expected from the model. 
\label{fl}}
\end{figure}



\begin{thebibliography}{}
\bibitem[Bohren \& Hufmann(1983)]{bohren83} Bohren, C.~F., \& Huffman, D.~R.\ 1983, 
Absorption and Scattering of Light by Small Particles (New York: Wiley)
\bibitem[Bottke et al.(2007)]{bottke07} Bottke, W.~F., 
Vokrouhlick{\'y}, D., \& Nesvorn{\'y}, D.\ 2007, \nat, 449, 48 
\bibitem[Dorschner et al.(1995)]{dorschner95} Dorschner, J., Begemann, 
B., Henning, T., Jaeger, C., \& Mutschke, H.\ 1995, \aap, 300, 503 
\bibitem[Fabian et al.(2000)]{fabian00} Fabian, D., J{\"a}ger, C., Henning, 
T., Dorschner, J., \& Mutschke, H.\ 2000, \aap, 364, 282 
\bibitem[Fabian et 
al.(2001)]{fabian01} Fabian, D., Henning, T., J{\"a}ger, C., Mutschke, H., Dorschner, J., 
\& Wehrhan, O.\ 2001, \aap, 378, 228 
\bibitem[Fujiwara et al.(2010)]{fujiwara10} Fujiwara, H., Onaka, 
T., Ishihara, D., et al.\ 2010, \apjl, 714, L152 
\bibitem[Hammel et al.(2010)]{hammel10} Hammel, H.~B., Wong, 
M.~H., Clarke, J.~T., et al.\ 2010, \apjl, 715, L150 
\bibitem[Holman 
\& Wiegert(1999)]{holman99} Holman, M.~J., \& Wiegert, P.~A.\ 1999, \aj, 117, 621 
\bibitem[Holmberg et 
al.(2009)]{holmberg09} Holmberg, J., Nordstr{\"o}m, B., \& Andersen, J.\ 2009, \aap, 501, 941 
\bibitem[Houck et al.(2004)]{houck04} Houck, J.~R., et al.\ 2004, \apjs, 154, 18 
\bibitem[Ishiguro et al.(2011)]{ishiguro11a} Ishiguro, M., 
Hanayama, H., Hasegawa, S., et al.\ 2011, \apjl, 740, L11 
\bibitem[Ishihara et al.(2010)]{ishihara10} Ishihara, D., Onaka, T., Kataza, H., et al.\ 2010, \aap, 514, A1 
\bibitem[Kemper et al.(2004)]{kemper04} Kemper, F., Vriend, 
W.~J., \& Tielens, A.~G.~G.~M.\ 2004, \apj, 609, 826 
\bibitem[Kenyon 
\& Bromley(2004)]{kenyon04} Kenyon, S.~J., \& Bromley, B.~C.\ 2004, \apjl, 602, L133 
\bibitem[Koike et al.(1989)]{koike89} Koike, C., Komatuzaki, T., Hasegawa, H., 
\& Asada, N.\ 1989, \mnras, 239, 127 
\bibitem[Kurucz(1992)]{kurucz92} Kurucz, R.~L.\ 1992, The 
Stellar Populations of Galaxies, 149, 225 
\bibitem[Lisse et al.(2009)]{lisse09} Lisse, C.~M., Chen, 
C.~H., Wyatt, M.~C., Morlok, A., Song, I., Bryden, G., 
\& Sheehan, P.\ 2009, \apj, 701, 2019 
\bibitem[Melis et al.(2010)]{melis10} Melis, C., Zuckerman, B., Rhee, J.~H., \& Song, I.\ 2010, \apjl, 717, L57 
\bibitem[Mikouchi et al.(2007)]{mikouchi07} Mikouchi, T., 
Tachikawa, O., Hagiya, K., et al.\ 2007, Lunar and Planetary Institute 
Science Conference Abstracts, 38, 1946 
\bibitem[Murakami et al.(2007)]{murakami07} Murakami, H., et al.\ 
2007, \pasj, 59, 369 
\bibitem[Murray 
\& Thompson(1990)]{murray90} Murray, C.~D., \& Thompson, R.~P.\ 1990, \nat, 348, 499 
\bibitem[Onaka et al.(2007)]{onaka07} Onaka, T., et al.\ 2007, \pasj, 59, 401 
\bibitem[Oudmaijer et al.(1992)]{oudmaijer92} Oudmaijer, R.~D., van der Veen, W.~E.~C.~J., 
Waters, L.~B.~F.~M., Trams, N.~R., Waelkens, C., \& Engelsman, E.\ 1992, \aaps, 96, 625 
\bibitem[Pahlevan 
\& Stevenson(2007)]{pahlevan07} Pahlevan, K., \& Stevenson, D.~J.\ 2007, Earth and Planetary Science Letters, 262, 438 
\bibitem[Reach et al.(1995)]{reach95} Reach, W.~T., et al.\ 
1995, \nat, 374, 521 
\bibitem[Rhee et al.(2008)]{rhee08} Rhee, J.~H., Song, I., 
\& Zuckerman, B.\ 2008, \apj, 675, 777 
\bibitem[Sargent et al.(2009)]{sargent09} Sargent, B.~A., et al.\ 2009, \apj, 690, 1193 
\bibitem[Spahn 
\& Sponholz(1989)]{spahn89} Spahn, F., \& Sponholz, H.\ 1989, \nat, 339, 607 
\bibitem[Song et al.(2005)]{song05} Song, I., Zuckerman, B., 
Weinberger, A.~J., \& Becklin, E.~E.\ 2005, \nat, 436, 363 
\bibitem[Takeda(2007)]{takeda07} Takeda, Y.\ 2007, \pasj, 59, 
335 
\bibitem[van Leeuwen(2007)]{vanleeuwen07} van Leeuwen, F.\ 2007, 
Astrophysics and Space Science Library, 350,  
\bibitem[Wyatt et al.(2007)]{wyatt07} Wyatt, M.~C., Smith, R., 
Greaves, J.~S., Beichman, C.~A., Bryden, G., \& Lisse, C.~M.\ 2007, \apj, 658, 569 
\bibitem[Wyatt(2008)]{wyatt08} Wyatt, M.~C.\ 2008, \araa, 46, 339 
\end{thebibliography}
\end{document}